\documentclass[aip,apl,twocolumn,amsmath,amssymb,bibnotes]{revtex4}

\usepackage{amsmath}
\usepackage{mathtools}
\usepackage[english]{babel}
\usepackage[utf8]{inputenc}
\usepackage{hyperref}
\usepackage{textcomp} 
\usepackage{physics}
\usepackage{float}
\usepackage{dsfont}
\usepackage{enumitem}
\usepackage{MnSymbol}
\usepackage{comment}
\usepackage{color}

\newcommand{\ri}{\mathrm{i}}

\newcommand{\re}{{\rm e}}

\begin{document}

\title{Thermal radiation and near-field thermal imaging of a plasmonic Su-Schrieffer-Heeger chain}
	
\date{\today}
	
\author{Florian Herz and Svend-Age Biehs$^{*}$}
\affiliation{Institut f{\"u}r Physik, Carl von Ossietzky Universit{\"a}t, D-26111 Oldenburg, Germany}
\email{s.age.biehs@uni-oldenburg.de} 
	
\begin{abstract}
We investigate the direct thermal emission spectrum of a plasmonic Su-Schrieffer-Heeger chain of InSb nanoparticles as well as its far-field emission due to near-field scattering by a sharp tip as it is used in scattering type thermal microscopes. We discuss the measurability of the topological phase transition in such far-field and near-field experiments and highlight the signatures of the topological edge modes. We further discuss the impact of a substrate. 
\end{abstract}
\maketitle


Nanoscale thermal imaging microscopes like the thermal radiative scanning tunneling microscope (TRSTM)~\cite{DeWilde,Babuty}, the  thermal infrared near-field spectroscope (TINS)~\cite{Huth2011,Jones, OCallahan}, and the scanning noise microscope (SNoiM)~\cite{Lin,WengEtAl2018, Komiyama} are collecting far-field thermal radiation which is either scattered or directly emitted by a thermal profiler. The sharp tip at the foremost part of the thermal profiler allows to scan surfaces with a high lateral resolution. Interestingly, microscopes like the SNoiM that  scatter the evanescent near-field of the thermally fluctuating sources inside the sample also allow to measure local temperatures of these thermal sources like the electron temperature of a hot electron gas in a quantum well structure~\cite{WengEtAl2018}, for instance. Recently, another experimental technique has been introduced to measure thermal emission spectra of single silica spheres with sizes smaller than the thermal wavelength without any thermal profiler~\cite{ExperimentDeWilde}. 

First theoretical approaches have tried to model the signal obtained with the scattering type thermal microscopes by a single dipole representing the foremost part of the probe~\cite{Joulain2,Jarzembski,Herz2018,Herz2021}. A more realistic modeling of the tip apex~\cite{Edalatpour2} within the so-called discrete dipole approximation (DDA)~\cite{Yurkin} has been brought forward. These methods are based on a many-body treatment of thermal radiation within the framework of fluctuational electrodynamics~\cite{RMP} and have also been used to handle near-field radiation problems~\cite{Edalatpour1, Ekeroth, Villa}. Recently, the many-body theory for describing far-field emission of a collection of nanoscale emitters in vacuum~\cite{Ekeroth,Centini2015,Centini2020} has been generalized for arbitrary environments~\cite{Herz2022} where the temperatures of the environment and different emitters can be chosen separately. Within the DDA this method is flexible enough to calculate the direct thermal emission of macroscopic objects as measured in Ref.~\cite{ExperimentDeWilde} as well as the signal of different scattering based thermal microscopes for which the temperatures for the tip, the sample, and the environment are different.

\begin{figure}
	\includegraphics[width = 0.45\textwidth]{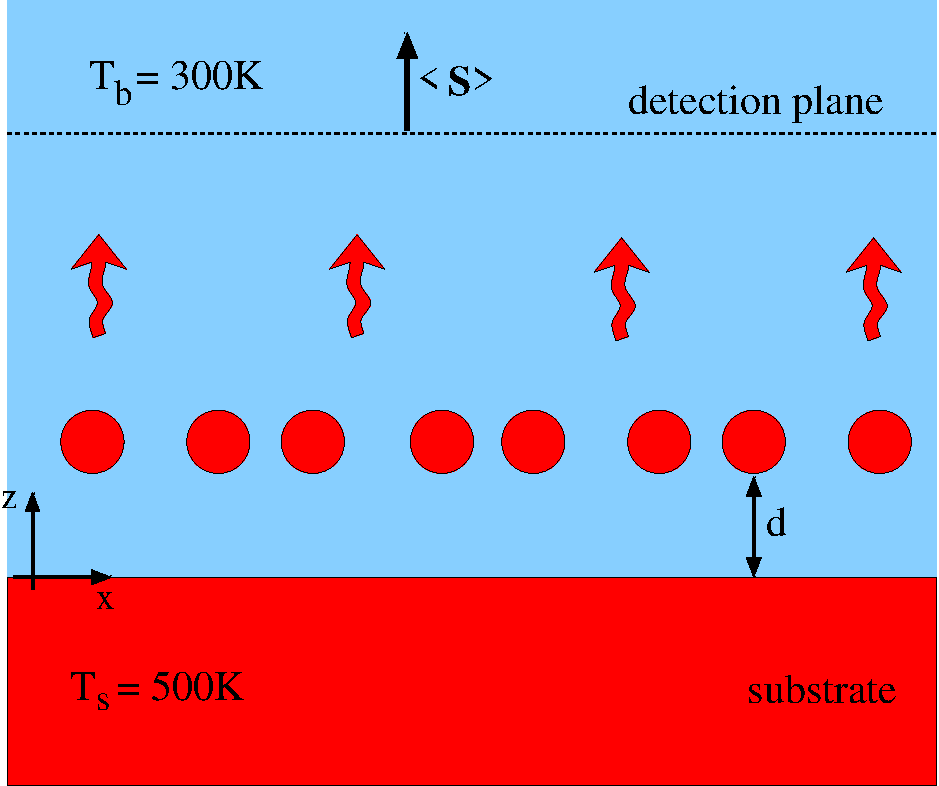}
	\caption{\label{Fig1} Sketch of a SSH NP chain with temperatures $T_1 = \ldots = T_N = T_{\rm ch}$ at distance $d$ to a substrate with temperature $T_s$ emitting thermal radiation into a vacuum background at temperature $T_b$. The emitted far-field power is determined by the integration of the mean Poynting vector $\langle \mathbf{S} \rangle$ over the observation plane indicated by the dashed line. The NPs of the two sub-lattices with lattice constant $a$ are labeled by $A$ and $B$. The distance between sub-lattices $A$ and $B$ is $t = \beta a/2$.}
\end{figure}

%
%
In this letter, we want to address theoretically the question whether it is possible to find a signature of the topological edge modes in a plasmonic Su-Schrieffer-Heeger (SSH) chain with the different kinds of scattering type near-field microscopes or the direct measuring method described in Ref.~\cite{ExperimentDeWilde}. That the SSH model can be realized with a relatively simple chain of plasmonic nanoparticles (NPs) has been shown by several theoretical works~\cite{Weick1,Weick2, Weick3, OESSH, ACSphotonSSH,JAPSSH}. Very recently, an experimental proof could be provided using a photo-emission electron microscopy~\cite{ExpSSH} technique. Since the topological edge modes can also be excited thermally and, therefore, contribute to the near-field radiative heat flux~\cite{OTTSSH,OTTHoney} and the thermal near-field energy density~\cite{OTTSSH2D}, we expect that also thermal radiation measurements will be able to measure the topological transition and maybe even the breakdown of the bulk-edge correspondence~\cite{PocockEtAl2019,OTTHoney} in such simple plasmonic SSH structures. To verify at least theoretically whether this expectation can be fulfilled with a direct thermal emission measurement as in Ref.~\cite{ExperimentDeWilde} or one of the scattering type near-field thermal microscopes \cite{DeWilde,Babuty,Huth2011,Jones, OCallahan,Lin,WengEtAl2018, Komiyama}, we calculate the directly far-field emitted spectral power $P_{\rm dir}$ of a collection of $N$ InSb NPs with temperatures $T_1 = \ldots = T_N \equiv T_{\rm ch}$ in a SSH chain configuration embedded in a vacuum background filled with a photon gas with temperature $T_b < T_{\rm ch}$ using the expression derived in Eq.~(62) of our work~\cite{Herz2022}. This expression can be obtained by calculating the mean Poynting vector $\langle \mathbf{S} \rangle$ due to thermal sources within the NPs only employing the framework of fluctuational electrodynamics. Then integrating it over an ``observation plane'' above the NP chain as depicted in Fig.~\ref{Fig1}
\begin{equation}
	P_{\rm tot} = \int  \!\!\! {\rm d} x \!\! \int \!\!\! {\rm d} y \, \langle \mathbf{S} \rangle \cdot \mathbf{e}_z 
\end{equation}
with the unit vector $\mathbf{e}_z$ in z-direction. This quantity gives then the emitted power by thermal radiation of the NP in the far-field $P_{\rm tot} = P_{\rm dir}$. All the details of the calculation can be found in Ref.~\cite{Herz2022}. By adding a substrate there will also be thermal sources in the substrate with temperature $T_s$ resulting in a contribution of thermal radiation $P_{\rm s}$ of the substrate itself. Additionally, one has to consider the part $P_{\rm abs}$ of the substrate emission absorbed within the NPs which does not reach the detection plane, a scattering part $P_{\rm sc}$ due to scattering between particles and substrate, and the direct part $P_{\rm dir}$ resulting in $P_{\rm tot} = P_{\rm s} + P_{\rm abs} + P_{\rm sc} + P_{\rm dir}$. The thermal emission of the substrate itself as well as the part of this thermal emission which is absorbed in the NPs will be omitted in the entire manuscript because we focus on the thermal emission of the NP chain itself. The explicit expression for the scattering part $P_{\rm sc}$ used below is given in Eq.~(68) of Ref.~\cite{Herz2022}. Finally, we also calculate the far-field scattering of a silicon tip as in the TINS experiments using the tip configuration from Ref.~\cite{Herz2022}. The results for $P_{\rm dir}$ and $P_{\rm sc}$ will allow us to explore theoretically the thermal emission spectra of SSH chains directly as it can be measured with direct methods like in Ref.~\cite{ExperimentDeWilde} or with the tip-based methods such as TRSTM, TINS, or SNoiM. 

%
%
Before we start to carry out the numerical evaluations, we revisit some properties of plasmonic SSH chains.
The eigenmodes in a chain of $N$ identical spherical NPs with radius $R$ described by a scalar polarizability $\alpha$ can be obtained following the procedure detailed in Ref.~\cite{ford}. The induced dipole moment $\mathbf{p}_i$ for a given NP $i$ ($i = 1, \ldots, N$) at a position $\mathbf{r}_i$ due to the field of all other dipoles is
$\mathbf{p}_i = \epsilon_0 \alpha \mathbf{E}(\mathbf{r}_i)$
with the field of all other dipoles given by
\begin{equation}
	\mathbf{E}(\mathbf{r}) = \mu_0 \omega^2 \sum_{j \neq i} \mathds{G}(\mathbf{r}_i, \mathbf{r}_j) \mathbf{p}_j.
\end{equation}
Here we make use of the Green's tensor $\mathds{G}(\mathbf{r}_i, \mathbf{r}_j)$. It can in general be written as the sum of a vacuum contribution $\mathds{G}_{\rm vac}$ and a scattering part $\mathds{G}_{\rm sc}$ which is important when considering a certain environment like a substrate. Furthermore, we have introduced the permittivity and permeability of vacuum $\epsilon_0$ and $\mu_0$ which are related to the vacuum light velocity by $c^2 = 1/\epsilon_0 \mu_0$.  Combining both equations into a block matrix equation gives the eigenvalue equation
\begin{equation}
  \tilde{\underline{\mathds{M}}} \underline{\mathbf{p}} = \frac{1}{\alpha} \underline{\mathbf{p}}
\label{Eq:EigenvalueEq}
\end{equation}
where the $3N$-dimensional block vector is defined by $\underline{\mathbf{p}} = (\mathbf{p}_1, \ldots, \mathbf{p}_N)^t$ and the $3N\times3N$ block matrix by ($i,j = 1, \ldots, N$)
\begin{equation}
	\tilde{\underline{\mathds{M}}}_{ij} = (1 - \delta_{ij})\frac{\omega^2}{c^2} \mathds{G}(\mathbf{r}_i, \mathbf{r}_j).
\end{equation}
For any given set of dipoles, the complex eigenfrequencies of the modes of the system can be determined by solving the eigenvalue equation (\ref{Eq:EigenvalueEq}) by determining the non-trivial solutions from the condition $\det(\tilde{\underline{M}} - \frac{1}{\alpha}\underline{\mathds{1}}) = 0$. In this work, we use the so-called strong-scattering expression of the polarizability which is given by
\begin{equation}
	\alpha_{\perp/\parallel} = \frac{V \chi_E}{1 - \frac{\omega^2}{c^2} V \chi_E \langle G_{\perp/\parallel} \rangle}
\end{equation}
where $\chi_E = \epsilon - 1$ is the susceptibility of the NP with permittivity $\epsilon$ and $V = 4 \pi R^3 / 3$ is its volume. The volume average of the Green's function $\langle \mathds{G} \rangle = {\rm diag}(\langle G_{\perp} \rangle, \langle G_{\perp} \rangle, \langle G_{\parallel} \rangle)$ is performed over the particle volume taking the full Green's function into account. It regularizes the divergence typically encountered when taking the self interaction into account. The explicit expressions for the case of a vacuum environment or a planar substrate can be found in Ref.~\cite{Herz2022}, for instance. 

\begin{figure}
	\includegraphics[width = 0.45\textwidth]{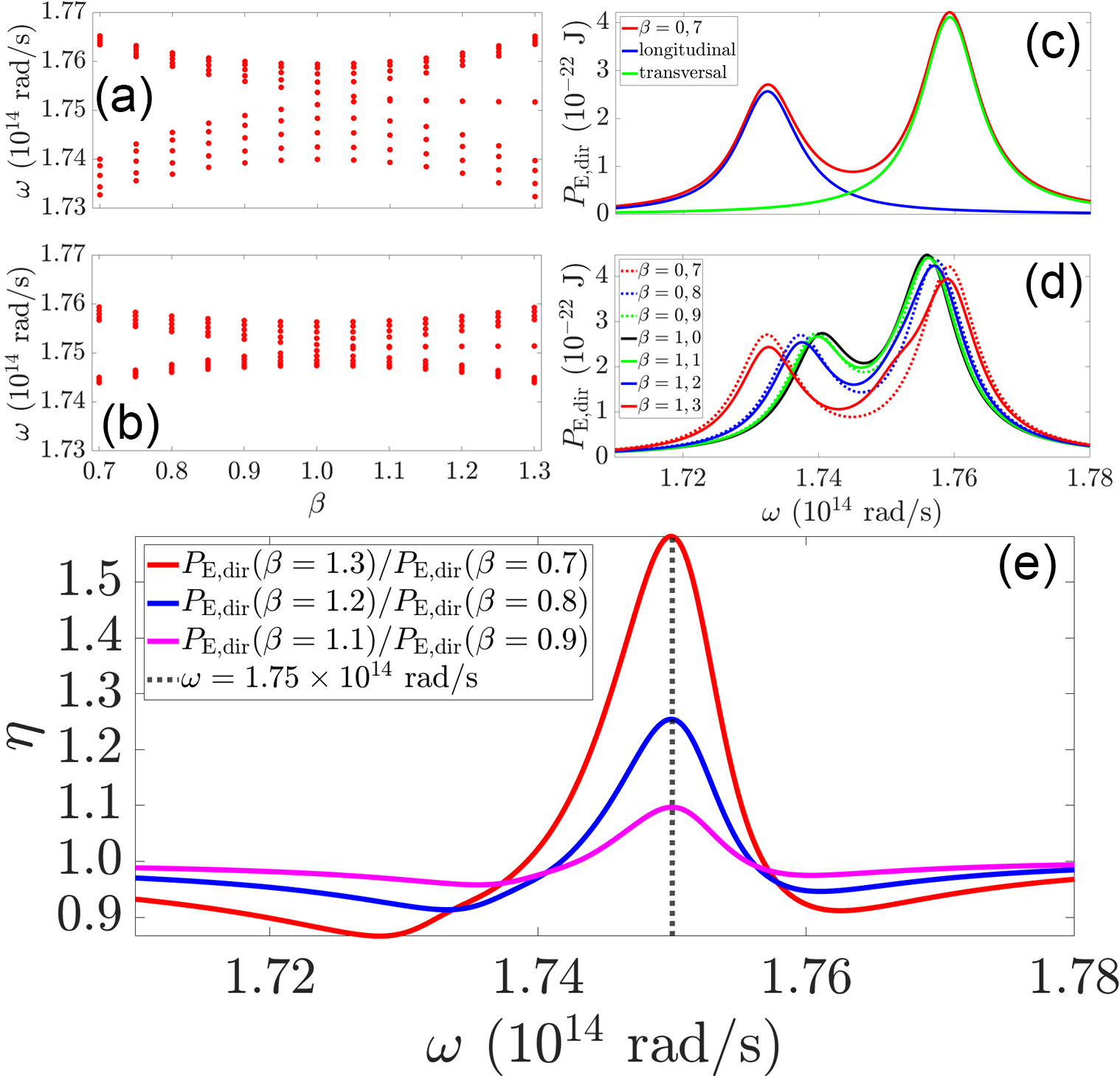}
	\caption{\label{Fig2} Thermal emission of a free standing SSH chain of $N = 10$ InSb NPs of radius $R = 100\,{\rm nm}$ and lattice constant $a = 8R$ assuming a temperature of $T_{\rm ch} = 700\,{\rm K}$ of the NPs and a temperature of $T_b = 300\,{\rm K}$ for the environment. (a) and (b) eigenmode frequencies for the longitudinal and transversal modes as a function of $\beta$ derived from the condition $\det(\tilde{\underline{M}} - \frac{1}{\alpha}\underline{\mathds{1}}) = 0$. (c) Thermal emission spectrum $P_{\rm dir,\omega}$ for different values of $\beta$. (d) Thermal emission spectrum $P_{\rm dir,\omega}$ for $\beta = 0.7$ showing the contribution of the longitudinal and transversal modes, separately.  (e) Ratio $\eta = P^{\beta = 1 + x}/P^{\beta = 1 - x}$ for $x = 0.1, 0.2, 0.3$. The vertical dashed line indicates the single particle resonance frequency $\omega = 1.75\times10^{14}\,{\rm rad/s}$.}
\end{figure}

Interestingly, when considering an infinite chain of NPs along an axis parallel to the x-axis with two particles per unit cell labeled by $A$ and $B$ as depicted in Fig.~\ref{Fig1} and introducing a Bloch ansatz for the dipole moments the eigenvalue equation can be translated to a $6$ dimensional eigenvalue equation
\begin{equation}
	\underline{\mathds{M}} \underline{\mathbf{p}}_{AB} = \frac{1}{\alpha} \underline{\mathbf{p}}_{AB}
\end{equation}
relating the eigenfrequencies to the Bloch vector $k$. Now, we have the block vectors $\underline{\mathbf{p}}_{AB} = (\mathbf{p}_A, \mathbf{p}_B)^t$ for the dipole moments of the two particles $A$ and $B$ in the unit cell and the block matrix ($\gamma, \delta = A,B$)
\begin{equation}
	\underline{\mathds{M}}_{\gamma,\delta} = \frac{\omega^2}{c^2} \sum_{j \neq 0} \mathds{G}(\mathbf{r}_{\gamma_0}, \mathbf{r}_{\delta_j}) \re^{\ri k x_{\delta_j}}
\end{equation}
where the positions of the particles are defined by $\mathbf{r}_{A_j} = (j d, 0, z)^t$ and $\mathbf{r}_{B_j} = ((j + \beta/2) d, 0, z)^t$ where $x_{\delta_j}$ is the corresponding x-component of $\mathbf{r}_{A_j}$ or $\mathbf{r}_{B_j}$. Note that we shifted the particle positions by $z$ from the x-axis because we will later consider the chain at a distance $z = d + R$ above a substrate which will be in the x-y plane as indicated in Fig.~\ref{Fig1}. If there is no substrate and the chain is in vacuum, then the Green's tensor is diagonal $\mathds{G} = \mathds{G}_{\rm vac} = {\rm diag}(G_{\parallel}, G_{\perp}, G_{\perp})$ due to the high symmetry with the explicit expressions for the components given in Ref.~\cite{OTTSSH}, for instance. In this case, the $6$ dimensional eigenvalue equation splits into a two dimensional eigenvalue equation for the longitudinal modes and two identical eigenvalue equations for the transversal modes in the chain ($\nu = \parallel, \perp$)
\begin{equation}
    \underline{\mathds{M}}_\nu \underline{\mathbf{p}}_{AB, \nu} = \frac{1}{\alpha} \underline{\mathbf{p}}_{AB, \nu}
\label{Eq:EigenvalueSSH}
\end{equation}
where $\underline{\mathbf{p}}_{AB,\nu} = (p_{A,\nu}, p_{B, \nu})^t$ contains only the components of the dipoles parallel or perpendicular to the chain axis (here, either $x$ or $y$/$z$ direction) and correspondingly ($\gamma, \delta = A,B$)
\begin{equation}
  \underline{M}_{\gamma,\delta, \nu} = \frac{\omega^2}{c^2} \sum_{j \neq 0} G_\nu(\mathbf{r}_{\gamma_0}, \mathbf{r}_{\delta_j}) \re^{\ri k x_{\delta_j}}.
\end{equation}
It has now been shown that exactly the eigenvalue equations (\ref{Eq:EigenvalueSSH}) are in the quasi-static limit equivalent to a two-band Schr\"{o}dinger equation for the SSH model when focusing to nearest-neighbor interaction~\cite{Weick1,Weick2, Weick3, OESSH, ACSphotonSSH,JAPSSH}. As a consequence, in this limit it can be shown that the Zak phase is either $0$ for $\beta \leq 1$ or $\pi$ for $\beta > 1$ so that there is a phase transition from a topological trivial phase (TTP) for $\beta \leq 1$ to a non-trivial phase (TNTP) for $\beta > 1$. This transition is robust when including long-range interactions, retardation, and dissipation~\cite{Weick2,Weick3},

%
%

Now, the bulk edge correspondence guarantees that in a finite chain of NPs there will be topologically protected edge modes in exactly that TNTP. These edge modes have been studied for plasmonic SSH chains in great detail~\cite{Weick1,Weick2, Weick3, OESSH, ACSphotonSSH,JAPSSH}. They can increase the near-field energy density of the particles at the edge of the chain~\cite{OTTSSH2D}, but also provide an important heat flux channel along the chain~\cite{OTTSSH}. In Fig.~\ref{Fig2}(a) and (b) we show the eigenmode frequencies for the longitudinal and transversal modes in vacuum for different values of $\beta$ for a chain of InSb NPs. It can first be seen that there is a symmetry for $\beta = 1 \pm x$ for different values of $x$. This symmetry is only broken by the appearance of the edge modes in the band gap for $\beta > 1$. When looking at the far-field spectrum of the direct thermal emission of such a chain of NPs in Fig.~\ref{Fig2}(c) and (d) we can see that the thermal emission is given by a contribution of the lower band of the longitudinal modes and a contribution of the upper band of the transversal modes. This is because only the modes in these bands have a dipole moment per unit cell and are, therefore, also called bright modes in contrast to the dark modes~\cite{Weick1} in the other bands which do not couple to the radiation field. Apart from this feature, it can be observed that the topological edge modes do contribute to the thermal emission even though these modes are highly localized at the edges of the chains~\cite{Weick1,OESSH,ACSphotonSSH,JAPSSH}. To visualize this feature, we plot the emission ration
\begin{equation}
  \eta = \frac{ P^{\beta = 1 + x}}{P^{\beta = 1 - x}}
\label{Eq:ratioeta}
\end{equation}
for  $x = 0.1, 0.2, 0.3$. In Fig.~\ref{Fig2}(e) it can be seen that the far field emission patterns for $\beta = 1 \pm x$ are similar but there is a noticeable difference between the spectra $\beta > 1$ and $\beta < 1$ at the edge mode frequency $\omega_{\rm em} = 1.751\times10^{14}\,{\rm rad/s}$ which approximately coincides with the single particle resonance frequency  $\omega = 1.75\times10^{14}\,{\rm rad/s}$. This feature indicates a stronger far-field emission at the edge mode frequency in the TNTP. On the other hand, for $\beta > 1$ the emission of the two bands is slightly diminished which is in agreement with the fact that one mode of each band for $\beta < 1$ becomes an edge mode so that for $\beta > 1$ less band-modes can radiate into the far-field for $\beta > 1$. 

\begin{figure}
	\includegraphics[width = 0.45\textwidth]{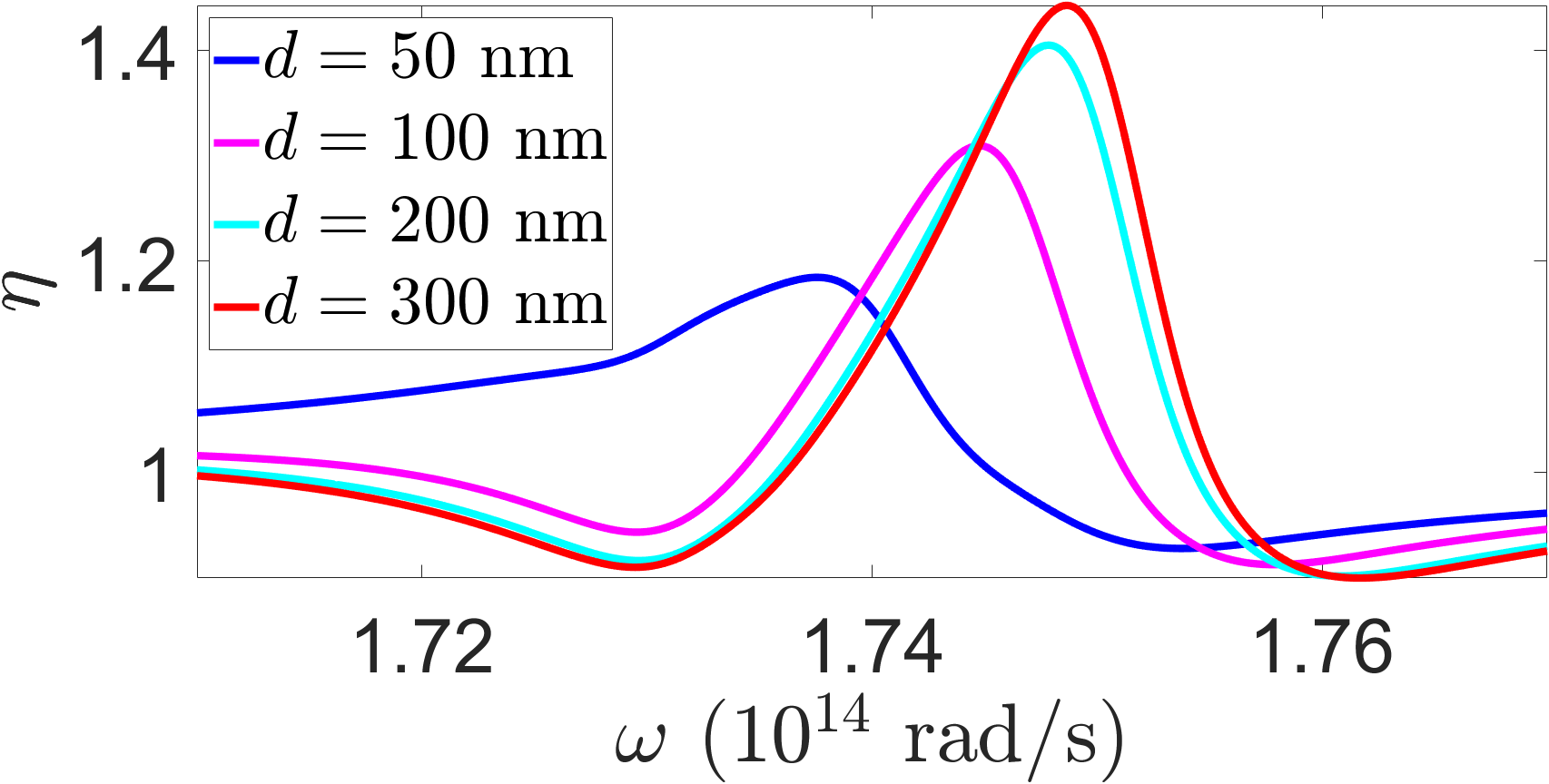}
	\caption{\label{Fig3} Thermal emission of the SSH chain with parameters as in Fig~\ref{Fig2} in edge to edge distance $d = 50\,{\rm nm}, 100\,{\rm nm},  200\,{\rm nm},  300\,{\rm nm}$ to a Ge substrate. Ratio $\eta = P^{\beta = 1.3}/P^{\beta = 0.7}$ from Eq.~(\ref{Eq:ratioeta}) for the sum $P = P_{\rm dir} + P_{\rm sc}$ of direct thermal emission of the chain $P_{\rm dir}$ and the scattering part $P_{\rm sc}$. We choose $T_s = T_{\rm ch} = 500\,{\rm K}$ and $T_b = 300\,{\rm K}$. }
\end{figure}

Let us now turn to the impact of a substrate. As substrate we choose Ge which basically has no dispersion or dissipation but a relatively large permittivity of $\epsilon_{\rm Ge} = 16$ in the infrared region. Now, we have not only a direct thermal emission $P_{\rm dir}$ of the NP chain but, due to the presence of the substrate, also a scattered part $P_{\rm sc}$, direct thermal emission of the substrate $P_{\rm sub}$, and thermal emission of the substrate $P_{\rm abs}$ which is absorbed by the particles. Here, we only focus on the scattering part and the direct thermal emission of the NPs because these parts contain the information about the thermal emission of the NPs. In Fig.~\ref{Fig3} the features are very similar to the SSH chain without substrate in Fig.~\ref{Fig2}. The difference is that, due to the presence of the surface, we obtain two edge modes; a transversal edge mode (y polarized) and a mixed transversal-longitudinal mode (z-x polarized). In Fig.~\ref{Fig3} we only see the enhanced $\eta$ due to the mixed edge mode. The edge mode resonance is more and more red-shifted when approaching the substrate as could be expected from previous observations~\cite{Babuty,Edalatpour2,Herz2022}. For example, for $d = 100\,{\rm nm}$ we find the resonance at $\omega = 1.745\times10^{14}\,{\rm rad/s}$ which is slightly smaller than the edge mode's eigenfrequency $\omega_{\rm em,mix} = 1.747\times10^{14}\,{\rm rad/s}$. Such small differences between the position of the resonance in $\eta$ and the exact edge mode frequency has already been observed for the free chain in Fig.~\ref{Fig2}. Furthermore, the main contribution stems from the directly emitted part of the NPs. We want to emphasize that the curves for $50\,{\rm nm}$ have to be taken with some caution because for such small distances multi-polar modes might play an important role. Therefore, to obtain results for the case where the NPs touch the interface, more elaborated DDA calculations for each NP of the chain have to be carried out which are out of the scope of this letter. However, the trend of the red-shift of the edge mode resonance seems reasonable, since it approaches the single particle resonance frequency observed in Fig.~\ref{Fig2}(e) without substrate. Hence, even by placing the NPs on a substrate, the edge modes will lead to an increased thermal emission at the corresponding red-shifted edge mode frequency. Therefore, far-field measurements like in Ref.~\cite{ExperimentDeWilde} could be carried out to observe the topological phase transition in an SSH chain of NPs.

%
%

\begin{figure}
	\includegraphics[width = 0.5\textwidth]{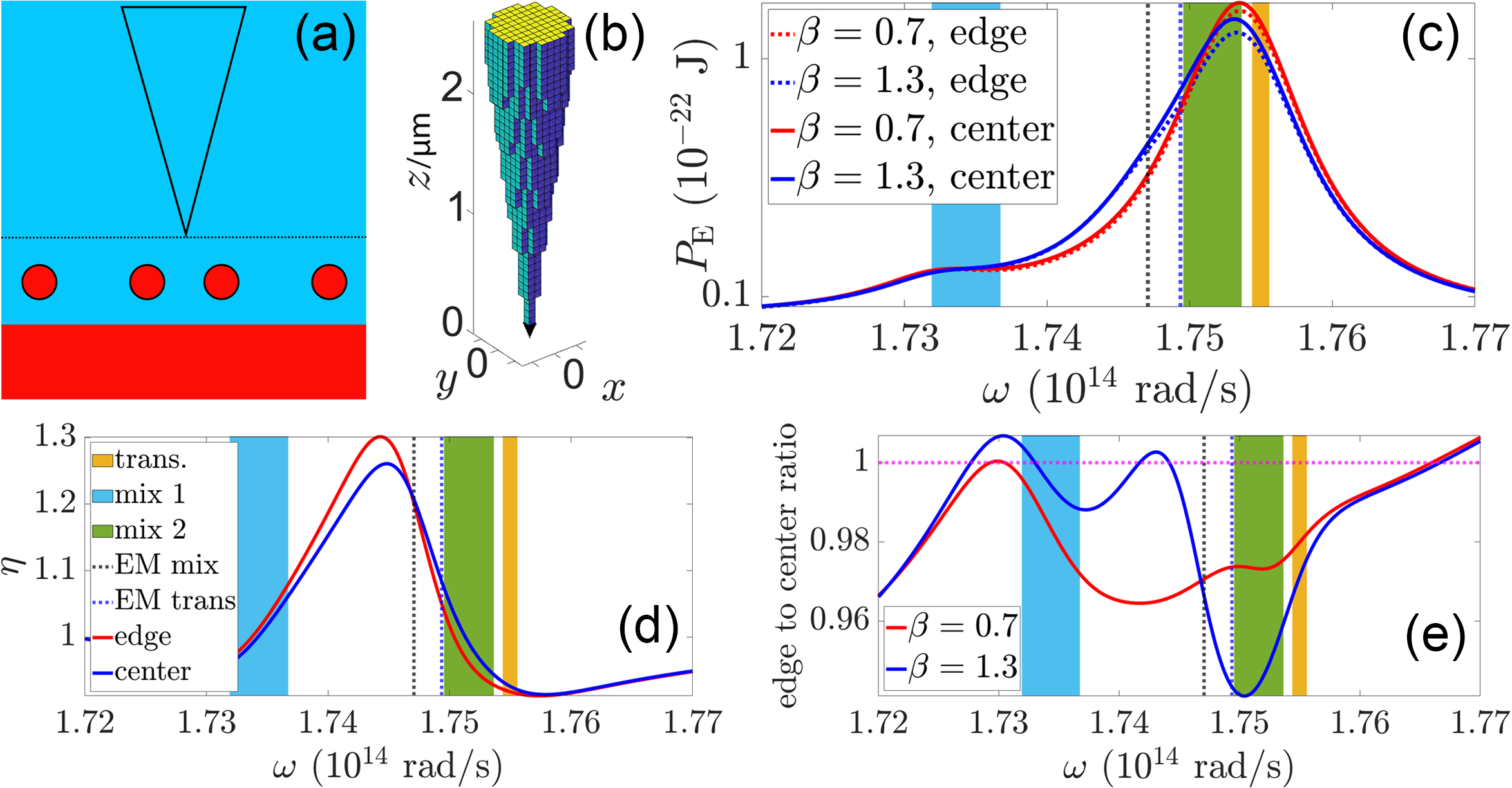}
	\caption{\label{Fig4} Thermal emission and scattering of the SSH chain of $N = 10$ InSb NPs in edge to edge distance $d = 100\,{\rm nm}$ to a Ge substrate due to the presence of a scattering tip. (a) Sketch of the configuration. (b) DDA of the scattering tip made of 5388 Si voxels. (c) We show the total emitted spectral power $P = P_{\rm dir} + P_{\rm sc}$ for the tip over an edge and a center particle for $\beta = 0.7$ and $1.3$. (d) The ratio $\eta = P^{\beta = 1.3}/P^{\beta = 0.7}$ from Eq.~(\ref{Eq:ratioeta}) for $P = P_{\rm dir} + P_{\rm sc}$. (e) The ratio of $P$ above the edge particle and $P$ above the center particle for $\beta = 0.7$ and $1.3$. We choose $T_s = T_{\rm ch} = 500\,{\rm K}$ and $T_{\rm tip} = T_b = 300\,{\rm K}$. The shaded areas visualize the bright bands of the mixed and transversal modes and the vertical dashed lines the corresponding edge mode frequencies $\omega_{\rm em,mix} = 1.747\times10^{14}\,{\rm rad/s}$ and $\omega_{\rm em, tr} = 1.749\times10^{14}\,{\rm rad/s}$.}
\end{figure}

Finally, we want to examine the possibility to measure the topological phase transition in the SSH chain with a thermal profiler scanning at an edge-to-edge distance of 50 nm above the SSH chain as depicted in Fig.~\ref{Fig4}(a). To this end, we consider a Si tip modeled by $5338$ voxels within the DDA as in Ref.~\cite{Herz2022} and as depicted in Fig.~\ref{Fig4}(b). As already seen in Fig.~\ref{Fig3}, the bands and resonances shift due to presence of the substrate and, here, also due to the tip so that the total emitted power $P = P_{\rm dir} + P_{\rm sc}$ is similar to $P_{\rm dir}$ for the free chain in Fig.~\ref{Fig2}(c) and (d) but with a red-shift. As might be expected, $\eta$ shown in Fig.~\ref{Fig4}(d) is quite similarly enhanced at the edge mode frequency as in the case without tip shown in Fig.~\ref{Fig3} when placing the tip over the edge NP of the chain. Interestingly this feature also exists if the tip is positioned over the center NP even though the enhancement is slightly smaller for the center NP than for the edge NP. The most striking feature is shown in Fig.~\ref{Fig4}(e) where the ratio of $P_e/P_c$ is plotted. Here, $P_e$ is the emitted spectral power for the tip over the edge particle and $P_c$ is the spectral emitted power when having the tip over the center particle. First, it can be seen that in the regions of the lower and upper bands around $1.73\times10^{14}\,{\rm rad/s}$ and $1.755\times10^{14}\,{\rm rad/s}$ the ratio $P_e/P_c$ exhibits a peak for the TTP and TNTP. On the other hand, at the frequency  $1.743\times10^{14}\,{\rm rad/s}$ we see that $P_e/P_c$  is increased and shows a local maximum in the TNTP for $\beta = 1.3$ [as also seen in Fig.~\ref{Fig3} and \ref{Fig4}(d)], whereas for the TTP with $\beta = 0.7$ the ratio $P_e/P_c$ shows a local minimum. Furthermore, at the frequency $1.75\times10^{14}\,{\rm rad/s}$ we find the opposite behavior;  $P_e/P_c$ shows a local minimum for the TNTP and maximum for the TTP. We associate these features with the mixed transversal-longitudinal and the transversal edge modes at $\omega_{\rm em,mix} = 1.747\times10^{14}\,{\rm rad/s}$ and $\omega_{\rm em, tr} = 1.749\times10^{14}\,{\rm rad/s}$, respectively. Note that the transversal mode frequency coincides very well with the frequency where $P_e/P_c$ shows a local minimum, whereas the mixed edge mode frequency $1.747\times10^{14}\,{\rm rad/s}$ is a little bit higher than the frequency $1.743 \times10^{14}\,{\rm rad/s}$ for which we observed the maximum in  $P_e/P_c$ and for $\eta$ in Figs.~\ref{Fig3}. This shift of resonance and edge mode frequency is in agreement with the observation of the free SSH chain in Fig.~\ref{Fig2}(e) and the SSH chain above the Ge substrate in Fig.~\ref{Fig3}. 

%
%

In conclusion, we have first shown that the thermal far-field emission of a single SSH chain of NPs has a distinct contribution of the edge modes in the TNTP. Furthermore, the presence of a substrate leads to a red-shift of the photonic bands in the NP chain and the edge mode frequency. The increased thermal emission at the edge-mode frequency can still be observed so that the topological phase transition should be observable in far-field experiments like in Ref.~\cite{ExperimentDeWilde}. Furthermore, we have shown that with a thermal near-field profiler, such as in TRSTM~\cite{DeWilde,Babuty}, TINS~\cite{Huth2011,Jones, OCallahan}, and SNoiM~\cite{Lin,WengEtAl2018, Komiyama}, similar features can be seen as for the direct thermal emission of the SSH chain. However, the advantage of the thermal profiler is that it can measure locally and our results indicate that in the TNTP there is an enhancement of the thermal emission when the tip is over the edge particle compared to the case where the tip is over the center particle. This enhancement becomes striking when the topological trivial and non-trivial cases are compared so that there is a clear signature of the edge states measurable with a near field thermal profiler.


S.-A.\ B.\ acknowledges support from Heisenberg Programme of the Deutsche Forschungsgemeinschaft (DFG, German Research Foundation) under the project No.\ 404073166 and F.\ H.\ acknowledges support from the Studienstiftung des deutschen Volkes.

%
%
\appendix

\end{document}